%% file: main.tex
\documentclass[conference]{IEEEtran}
\usepackage[T1]{fontenc}
\usepackage{amsmath,amssymb,amsthm,booktabs,graphicx,hyperref}
\input{paper_title}
\hypersetup{hidelinks,pdfauthor={Shaoyang Sun and Chongwen Huang},pdftitle={\papertitle}}
\graphicspath{{figures/}}
\newtheorem{proposition}{Proposition}
\newcommand{\norm}[1]{\left\lVert#1\right\rVert}
\title{\papertitle}
\input{authors}
\begin{document}
\maketitle
\begin{abstract}
Pattern-reconfigurable arrays acquire channels by distributing radiation modes across their antenna elements. Equal mode counts fix the observation budget while leaving a spatial sampling choice that directly affects reconstruction of unmeasured channels. We establish how this assignment enters the inverse problem through the receive aperture and the coupling between propagation paths. For an isolated path with calibrated positive real responses, the power-weighted variance of antenna positions completely characterizes the assignment dependence of the profiled angle information at fixed mode counts. We then construct layouts with identical weighted apertures and different path correlations, exposing a limitation of aperture-based design in multipath propagation. A paired simulation study evaluates balanced random, contiguous, interleaved, and equal-moment sampling using a common estimator with per-detection joint angle refinement. Balanced random sampling substantially reduces complete-channel error relative to ascending contiguous blocks when some modes are entirely unobserved, with improvements across nearly all tested scenes at high signal-to-noise ratio. When every mode is sampled, the mean benefit is more sensitive to rare large errors. Together, the analysis and experiments connect the spatial distribution of mode measurements to both extrapolation accuracy and reconstruction stability.
\end{abstract}


\section{Introduction}
Pattern-reconfigurable MIMO arrays expose different channels by changing their radiation modes. Recovering all antenna--mode pairs from partial observations requires extrapolating their shared propagation structure. Even with a fixed observation budget and equal mode counts, the receiver can assign each mode to adjacent elements or distribute it across the aperture. These choices sample different spatial phases and can change how well the underlying paths are resolved.

Prior studies predict untrained channels from pattern correlations~\cite{Armandoust2022Joint,Dinis2026Pixel}, select modes for direction finding~\cite{Gurbuz2020Modes}, learn extrapolation from grouped measurements~\cite{Liang2024PRNet}, or estimate angles and gains using reference and grouped acquisitions~\cite{Liang2025LowPilot,Liang2026EfficientExtrapolation}. These approaches, together with spatial interpolation and sensor selection~\cite{Do2025Interpolation,VanderWerf2024Multiplicity}, motivate examining where a prescribed set of mode measurements should lie.

Our contributions connect this placement to the geometry and accuracy of calibrated channel reconstruction.
\begin{itemize}
\setlength\itemsep{1pt}
\item We derive the profiled angle-information matrix for an isolated path. At fixed mode counts, its assignment dependence is fully captured by the power-weighted position variance.
\item We express multipath coupling through assignment polynomials and construct layouts with identical aperture moments but different path correlations.
\item On 512 independent ten-path scenes, balanced random sampling reduces mean complete-channel NMSE by 46.8\% relative to ascending blocks when only two modes are measured, improving 95.5\% of scenes at 20 dB. Same-scene diagnostics examine the larger tail sensitivity when every mode is sampled.
\end{itemize}
All layouts use the same pursuit and joint-refinement estimator, allowing their spatial sampling effects to be compared directly.

\begin{figure*}[!t]
\centering
\includegraphics[width=\textwidth]{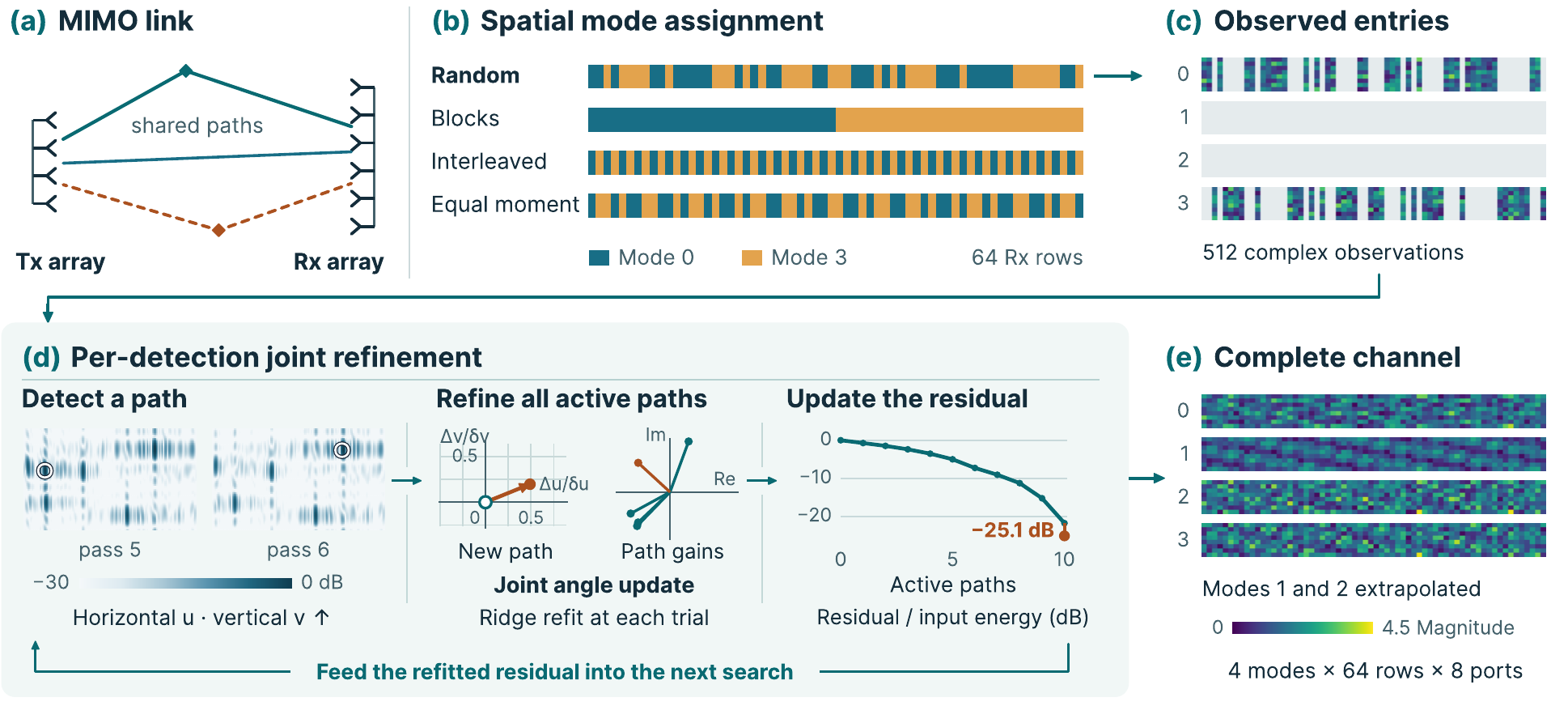}
\caption{\textbf{From spatially assigned measurements to the complete MIMO channel.} (a) Schematic arrays and shared paths. (b) Each layout assigns modes 0 and 3 to 32 receive elements each. (c) The balanced random layout gives 512 complex observations. Within each mode band, columns index receive rows and rows index transmit ports. Gray entries are unobserved. (d) Two successive residual-correlation maps from the first 20 dB scene (first noise draw). Each map uses $20\log_{10}$ correlation amplitude relative to its peak, with receive frequency $u$ horizontal and transmit frequency $v$ increasing upward. Circles mark selected peaks. At pass five, joint refinement moves the new path off grid (displacement in receive and transmit grid spacings) and refits all five complex gains. The orange endpoint marks the final joint fit at fixed order. (e) The fitted paths render all four modes, including the two unsampled modes. The observed and reconstructed magnitudes share the scale in (e).}
\label{fig:acquisition}
\end{figure*}

\section{Channel Model and Spatial Assignment}
We consider a narrowband link with half-wavelength uniform linear arrays. The receiver has $N_r=64$ elements and four calibrated radiation modes, and the transmitter has $N_t=8$ ports. The complete channel is
\begin{equation}
 H_{knt}=\sum_{\ell=1}^{L}c_\ell q_k(u_\ell)
           e^{j\pi n u_\ell-j\pi t v_\ell},
 \label{eq:model}
\end{equation}
where $k,n,t$ index the mode, receive element, and transmit port. The spatial frequencies are $u_\ell=\sin\phi_{r,\ell}$ and $v_\ell=\sin\phi_{t,\ell}$. The path gain $c_\ell$ and geometry are shared across modes, while $q_k(u_\ell)$ specifies the calibrated receive field response. With angles in degrees, we use
\begin{align}
 \psi_k(u)&=\operatorname{wrap}_{[-180,180)}
       \!\left(180\arcsin(u)/\pi+\delta_k\right),\nonumber\\
 q_k(u)&=10^{[8-\min\{12(\psi_k(u)/65)^2,30\}]/20},
 \label{eq:pattern}
\end{align}
with $\delta=(45,-45,15,-15)$ and $|u|\le\sin(\pi/3)$. Transmit frequency is periodic with period two. All indices start at zero.

Assignment $a_n$ specifies the mode measured at element $n$. Each receive element acquires all transmit ports,
\begin{equation}
 y_{nt}=H_{a_n,n,t}+e_{nt},\qquad
 e_{nt}\sim\mathcal{CN}(0,\sigma^2).
 \label{eq:observation}
\end{equation}
This provides $m=512$ complex observations for $2048$ target entries. The corresponding pilot-symbol count depends on the front end. We consider all four modes measured on 16 elements each and modes $\{0,3\}$ measured on 32 elements each. The latter requires extrapolating modes 1 and 2 across the entire array.

Four layouts preserve these counts. \emph{Balanced random} sampling uniformly permutes the assigned modes. \emph{Ascending blocks} allocate adjacent equal-size subarrays in increasing mode order. \emph{Interleaving} repeats the active-mode sequence. The \emph{equal-moment} construction gives every mode group the same position mean and variance. The estimator receives the observations, assignment, source and target response descriptors, and noise variance. The primary experiments supply the known order $L=10$. Figure~\ref{fig:acquisition} summarizes acquisition, path reconstruction, and rendering of the complete channel.

\section{Aperture Information and Multipath Coupling}
\subsection{Exact assignment dependence for an isolated path}
The calibrated responses connect each measured path to every target mode through~\eqref{eq:model}. Estimating its angle determines both the spatial phase progression and the mode-dependent amplitude. Spatial assignment therefore influences the information used for extrapolation even when a target mode is absent from the observations.

\begin{proposition}[Response-weighted aperture]
For an isolated path, assume an unconstrained complex gain $c\ne0$, positive real differentiable responses and a complete set of transmit ports at each receive element. Define
\[
 p_n=\frac{q_{a_n}(u)^2}{\sum_r q_{a_r}(u)^2},\quad
 \beta_n=\frac{q'_{a_n}(u)}{q_{a_n}(u)},\quad
 E=N_t\sum_n q_{a_n}(u)^2.
\]
After eliminating the unknown complex gain, the Fisher information matrix for $(u,v)$ is
\begin{equation}
 \frac{2|c|^2 E}{\sigma^2}\operatorname{diag}\!\left(
 \operatorname{Var}_p\beta+\pi^2\operatorname{Var}_p n,
 \frac{\pi^2(N_t^2-1)}{12}\right).
 \label{eq:aperture}
\end{equation}
\end{proposition}
\begin{proof}
Let $b$ denote the observed unit-gain path response. Its derivatives have entries $b_{nt}(\beta_n+j\pi n)$ and $-j\pi t b_{nt}$. Orthogonal projection off the complex span of $b$ centers $\beta_n,n,t$ under the weights $p_n/N_t$. The transmit index is then independent of the receive index, so the real cross term vanishes. The squared norms of the projected derivatives give~\eqref{eq:aperture}.
\end{proof}
At fixed mode counts and angle, $E$ and $\operatorname{Var}_p\beta$ are invariant. The power-weighted variance $\operatorname{Var}_p n$ therefore completely determines the assignment dependence in~\eqref{eq:aperture}. Larger weighted variance increases the receive-angle phase information.

\subsection{An aperture criterion and its multipath limitation}
Let $\alpha_k$ be the total normalized response power on mode $k$ and let $\mu_k,V_k$ be its group position mean and variance. Then
\begin{equation}
 \operatorname{Var}_p n=\sum_k\alpha_k V_k+
                         \operatorname{Var}_{\alpha}\mu_k .
 \label{eq:mixture}
\end{equation}
For balanced groups, uniform weights recover the full-array variance $341.25$. The minimum over positive group weights can therefore never exceed this value. Equal group means and variances attain it for every weight vector, maximizing this worst-case aperture criterion. With $K_s=2$ or 4 active modes, assigning element $n$ by the sum of its base-$K_s$ digits modulo $K_s$ yields such a partition~\cite{Bolker2016Prouhet}.

\begin{figure}[t]
\centering
\includegraphics[width=\columnwidth]{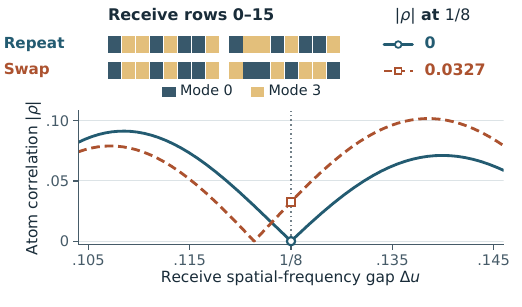}
\caption{\textbf{Equal aperture moments, different path correlations.} The strips show the first two eight-row blocks; the pattern extends to 64 rows. Repeat preserves the mode labels and Swap exchanges them in alternate blocks. For $u_1=0$ and equal transmit frequencies, $|\rho|=|b_1^H b_2|/(\norm{b_1}_2\norm{b_2}_2)$. At $\Delta u=1/8$, the correlations are zero and 0.0327.}
\label{fig:moments}
\end{figure}

In multipath propagation, the geometry fit also depends on the correlation between observed path responses. Set $P_k(z)=\sum_{n:a_n=k}z^n$ and $D_{N_t}(d)=\sum_{t=0}^{N_t-1}e^{-j\pi td}$. The inner product between the observed unit-gain responses of paths $\ell$ and $j$ is
\begin{equation}
 b_\ell^H b_j=D_{N_t}(v_j-v_\ell)
 \sum_k q_k(u_\ell)^*q_k(u_j)
 P_k\!\left(e^{j\pi(u_j-u_\ell)}\right).
 \label{eq:gram}
\end{equation}
Equal counts and position moments constrain each assignment polynomial and its first two derivatives at $z=1$. Distinct path frequencies instead sample its spectrum along the unit circle. As an explicit counterexample, divide every eight-element block into $\{0,3,5,6\}$ and its complement and assign modes 0 and 3. Repeating the labels in every block or exchanging them in alternate blocks preserves both group moments across the array. At $(u_1,u_2)=(0,1/8)$ and equal transmit frequencies, the normalized correlation magnitudes are zero and 0.0327, respectively. The correlation null shifts between layouts (Fig.~\ref{fig:moments}) despite identical single-path information. Matching aperture moments consequently leaves the multipath correlation structure undetermined.

\subsection{Local complete-channel sensitivity}
Let $Q$ span the real-valued complete-channel tangent with orthonormal columns, and let $S_R$ select observed coordinates. When $S_RQ$ has full column rank, local unregularized least squares has normalized noise risk
\begin{equation}
 R_{\mathrm{loc}}=\frac{\sigma^2}{2\norm H_F^2}
                   \sum_j\sigma_j(S_RQ)^{-2}.
 \label{eq:local}
\end{equation}
The factor $1/2$ follows from proper complex noise. The regular ten-path tangent has dimension 40 with free gains and 39 under their unit-norm constraint. This reference describes local perturbations around the true geometry. Nonlinear fits can return more distant solutions.

\section{Per-Detection Joint Reconstruction}
Per-detection joint refinement (PDJ) combines continuous-frequency pursuit~\cite{Mamandipoor2016NOMP,Zeng2024QNOMP} and regularized variable projection~\cite{GolubPereyra1973,OLearyRust2013} in a fixed reconstruction procedure. Its residual-driven loop adds a path, updates the joint geometry, and removes the fitted contribution before searching again.

\subsection{Factored detection using the measured array}
Let $R\in\mathbb C^{N_r\times N_t}$ contain the current residual, initialized to the measurements. For candidate transmit frequency $v$, first correlate across the complete transmit row. Then combine the rows using the receive steering phase and assigned mode response,
\begin{align}
 z_n(v)&=\sum_{t=0}^{N_t-1}R_{nt}e^{j\pi tv},\nonumber\\
 \gamma_a(u,v)&=
 \frac{\left|\sum_n q_{a_n}(u)^*e^{-j\pi nu}z_n(v)\right|^2}
 {N_t\sum_n|q_{a_n}(u)|^2}.
 \label{eq:detect}
\end{align}
The first operation aligns transmit phases. The second aligns receive phases and weights each element by its calibrated response. The denominator normalizes the observed atom energy, making scores comparable across directions with different radiation gains. Equation~\eqref{eq:detect} is exactly $|b_a(u,v)^H r|^2/\norm{b_a(u,v)}_2^2$ for vectorized residual $r$.

We append the unused grid pair maximizing $\gamma_a$ on a $128\times64$ receive--transmit grid. The row correlations $z_n(v_j)$ are reused across all receive candidates $u_i$. The receive and transmit factors require $64\times128$ and $8\times64$ complex entries, respectively. This exact factorization stores 8704 entries in place of a 4\,194\,304-entry dictionary.

\subsection{Joint refinement after eliminating the gains}
With $s$ active paths, $A(\theta)$ contains their observed atoms and $\theta=(u_1,\ldots,u_s,v_1,\ldots,v_s)$. The fitted objective is
\begin{align}
 Q_\lambda(\theta,c)&=m^{-1}\norm{A(\theta)c-y}_2^2+
                       \lambda\norm c_2^2,\label{eq:objective}\\
 c_\lambda(\theta)&=(A^H A+m\lambda I)^{-1}A^Hy,\label{eq:ridge}
\end{align}
where $\lambda=\max(10^{-10},0.05\sigma^2)$. Solving~\eqref{eq:ridge} at every candidate geometry removes $2s$ real gain variables from the nonlinear search. The remaining $2s$ frequencies minimize the squared norm of the augmented residual
\begin{equation}
 \zeta(\theta)=\frac1{\sqrt m}
 \begin{bmatrix}
 A(\theta)c_\lambda(\theta)-y\\
 \sqrt{m\lambda}\,c_\lambda(\theta)
 \end{bmatrix}.
 \label{eq:augmented}
\end{equation}
The lower block retains the coefficient penalty during angle updates. The trust-region solver uses the real and imaginary parts of $\zeta$ and its exact derivative through $c_\lambda$.

Cross-path correlations in~\eqref{eq:gram} explain why all active paths are updated. A newly detected direction changes the joint gain fit and the residual attributed to earlier paths. Refining only the new direction would hold their current angular errors fixed. Joint refinement lets the existing frequencies adjust before the next detection.

\subsection{Residual feedback and complete-channel rendering}
After refinement, set $R=\operatorname{reshape}(y-Ac_\lambda)$ and repeat~\eqref{eq:detect}. Each of the first $L-1$ updates permits five function evaluations. At $s=L$, evaluate the initial full-order fit and perform a final joint update capped at 40 evaluations. Substitution of the final paths into~\eqref{eq:model} renders every target mode, including those absent from acquisition.

The example in Fig.~\ref{fig:acquisition} follows scene 0, modes $\{0,3\}$, balanced random sampling, 20 dB, and noise repeat 0. Detection 5 selects $(-0.661,0.156)$, which jointly refines off grid to $(-0.655,0.162)$. Its updated residual drives detection 6. The next joint update also changes the second path's transmit frequency from $-0.945$ to $-0.952$, illustrating how adding a path updates earlier estimates. Relative residual energy $\norm R_F^2/\norm y_2^2$ falls from 0.444 to 0.316 and then 0.187 over these detections. The terminal call reaches 0.00308 and a full-channel NMSE of $1.80\times10^{-4}$. All layout comparisons use the same solver and budgets.

\section{Numerical Results}
\subsection{Matched acquisition and statistical protocol}
The estimator and analysis were fixed before drawing 512 independent scenes. Arrival and departure angles are uniform on $[-60,60]$ and $[-90,90]$ degrees. Ten proper complex Gaussian gains are jointly normalized to $\norm c_2=1$, with no minimum path separation or channel-energy normalization. Training-derived omnidirectional power $P_0=0.9997065$ sets $\sigma^2=P_0 10^{-\mathrm{SNR}/10}$. We test 0, 10, and 20 dB with two noise realizations. Layouts share the full target and noise field. Each scene-random layout is channel-independent and fixed across SNRs and repeats.

For scene $i$ and repeat $r$, define $e_{ir}=\norm{\widehat H_{ir}-H_i}_F^2/\norm{H_i}_F^2$. We average this normalized mean squared error (NMSE) in linear units over repeats and scenes. Our prespecified contrast at 20 dB is
\begin{equation}
 \Delta=10\log_{10}\!\left(\frac{\bar e_{\rm blocks}}
                                      {\bar e_{\rm random}}\right).
 \label{eq:effect}
\end{equation}
Positive values mean higher average error under ascending blocks. We drew 100\,000 paired bootstrap samples of whole scenes. The two primary comparisons use 97.5\% marginal percentile intervals for nominal 95\% familywise coverage under Bonferroni adjustment.

\begin{figure*}[t]
\centering
\includegraphics[width=\textwidth]{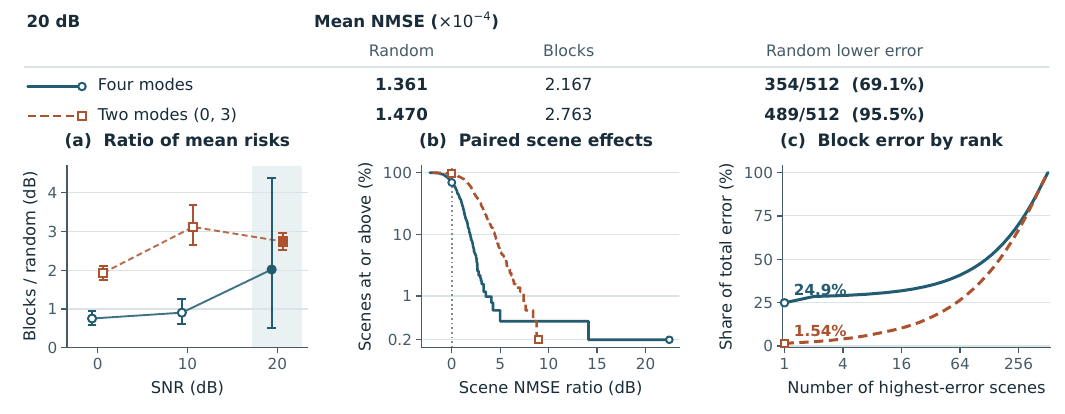}
\caption{\textbf{Matched assignment experiment on 512 scenes.} The top rows report mean NMSE and scene wins at 20 dB. (a) Blocks-to-random ratio of mean NMSE. Filled markers at 20 dB show the primary effects with 97.5\% marginal paired-bootstrap intervals; open markers show descriptive 95\% intervals. (b) Fraction of scenes at or above each within-scene ratio at 20 dB. (c) Cumulative share of total block error, ranked from the worst scene down, at 20 dB. Each scene averages two noise realizations in linear units.}
\label{fig:assignment}
\end{figure*}

\textbf{Extrapolation to unmeasured modes.}
With only modes 0 and 3 measured, mean NMSE at 20 dB was $1.47\times10^{-4}$ under balanced random assignment and $2.76\times10^{-4}$ under ascending blocks. Random sampling reduced mean error by \textbf{46.8\%}, corresponding to a block penalty of $\Delta=2.74$ dB (adjusted interval $[2.52,2.97]$ dB). It improved \textbf{489 of 512 scenes} (95.5\%) after averaging the two repeats. Entirely unmeasured modes account for 62.2\% of the excess linear error under blocks. All regional errors share the full-channel energy denominator.

\textbf{Partial observations of every mode.}
With all four modes measured, random assignment gave a mean NMSE of $1.36\times10^{-4}$, compared with $2.17\times10^{-4}$ for blocks. This gives a 37.2\% reduction and an effect of $2.02$ dB ($[0.52,4.39]$ dB). Random assignment improved 354 of 512 scenes (69.1\%). The largest-error scene accounted for 24.9\% of total block error, compared with 1.54\% in the two-mode experiment (Fig.~\ref{fig:assignment}). Omitting that scene from both layouts reduced the four-mode contrast to 0.785 dB. For the two-mode experiment, deleting any one scene gave a contrast between 2.69 and 2.75 dB. All primary results retain every scene.

At 20 dB, interleaving and the equal-moment layout had 0.12 and 0.11 dB lower mean error than random assignment with four acquired modes. With two acquired modes their descriptive intervals relative to random assignment both spanned zero. Their similar empirical risks contrast with the larger penalty for contiguous sampling in the two-mode setting.

\subsection{Local sensitivity, nonlinear fits, and block order}
\input{sections/en/mechanism}

\subsection{Estimator controls and calibration sensitivity}
On a separate 1000-scene cohort spanning $-10$ to 20 dB, PDJ outperformed grid-only, terminal-only, coordinate-only, independent-mode, and empirical linear minimum mean squared error (LMMSE) references in SNR-averaged mean NMSE. The adjusted comparison with C+J was inconclusive. A 128-scene CPU benchmark with one BLAS thread gave 45.8 ms per PDJ reconstruction in both acquisition settings, compared with 248.2/249.5 ms for C+J. The supplement specifies the baseline and timing protocols.

In a 128-scene calibration control, a common four-degree shift of the true patterns raised two-mode random-layout NMSE from $1.48\times10^{-4}$ to $4.13\times10^{-3}$ at 20 dB. Its block penalty fell from 3.11 to 0.85 dB. A smaller relative assignment effect can thus accompany substantially larger absolute error.

\section{Conclusion}
Spatial mode placement changes complete-channel reconstruction at a fixed measurement budget. The power-weighted aperture exactly captures its isolated-path information dependence, while equal-aperture layouts can have different multipath correlations. Balanced random sampling reduced two-mode mean NMSE by 46.8\% and improved 95.5\% of tested scenes at 20 dB. The four-mode benefit was more sensitive to rare errors and block order. These findings support spatially dispersed acquisition under the calibrated model. Measured complex responses and layout-dependent coupling remain to be evaluated.

\bibliographystyle{IEEEtran}\bibliography{references}
\end{document}

%% file: paper_title.tex
\newcommand{\papertitle}{The Geometry of Mode Sampling in Reconfigurable MIMO Channel Extrapolation}

%% file: authors.tex
\author{\IEEEauthorblockN{Shaoyang Sun and Chongwen Huang\textsuperscript{*}}
\IEEEauthorblockA{Zhejiang University, Hangzhou, China\\
3240105985@zju.edu.cn\qquad\textsuperscript{*}Corresponding author: chongwenhuang@zju.edu.cn}}

%% file: sections/en/mechanism.tex
Supplying the true angles to the ridge fit reduces the assignment effects to 0.07/0.17 dB for four/two measured modes (Table~\ref{tab:mechanism}). The local reference predicts 0.63/2.83 dB, matching the two-mode mean more closely. Its scene-level correlations with empirical ratios are 0.20/0.44, indicating substantial variation around the average relationship.

\begin{table}[t]
\caption{Contiguous-to-random ratios of mean NMSE at 20 dB on the same 512 scenes. The diagnostic rows use true path geometry.}
\label{tab:mechanism}\centering
\small\setlength{\tabcolsep}{4pt}\renewcommand{\arraystretch}{1.12}
\begin{tabular}{@{}lrr@{}}\toprule
 & \multicolumn{2}{c}{Assignment effect $\Delta$ (dB)}\\
\cmidrule(l){2-3}
Estimator or diagnostic & Four modes & Two modes\\\midrule
\multicolumn{3}{@{}l}{\textit{Reconstruction from observations}}\\
Per-detection joint refinement & \textbf{2.02} & \textbf{2.74}\\
Coordinate-plus-joint refinement & 1.90 & 2.85\\
\addlinespace[2pt]
\multicolumn{3}{@{}l}{\textit{At the true path geometry}}\\
Gain fit, observed noise & 0.07 & 0.17\\
Gain fit, expected over noise & 0.06 & 0.12\\
Local reference, free gains & 0.63 & 2.83\\
Local reference, unit-norm gains & 0.61 & 2.72\\\bottomrule
\end{tabular}
\end{table}

For the worst four-mode block scene, changing only the noise raises NMSE from $1.76\times10^{-4}$ to $5.51\times10^{-2}$. Three starts and a terminal cap of 200 retain objective 0.0290 on the difficult realization. A feasible true-angle gain fit attains 0.0107, proving the returned fit is suboptimal for its objective. Coordinate-plus-joint (C+J) refinement preserves the positive mean assignment effects in both settings.

A post-hoc enumeration of all contiguous block orders gives penalties of 0.23--12.37 dB across the 24 four-mode orders and 2.74--3.62 dB across the two-mode orders. Thus the four-mode result is sensitive to both rare fits and where each contiguous mode block is placed.